# ANÁLISIS DE LA PRECIPITACIÓN EN TLAQUEPAQUE: UNA MIRADA AL PASADO Y A LA ACTUALIDAD

*Mauricio López-Reyes[1, 2]*
*María Luisa Martín-Pérez[3]*
*Juan Jesús Gónzalez-Alemán[4]*
*[1]Instituto de Astronomía y Meteorología, Universidad de Guadalajara*
*[2] Departamento de Física de la Tierra y Astrofísica, Universidad Complutense de Madrid*
*[3]Departamento de Matemática Aplicada, Escuela de Ingeniería Informática, Universidad de Valladolid*
*[4]Agencia Estatal de Meteorología, Madrid, España*
*maurilop@ucm.es*

1. **Introducción**

Los eventos de precipitación que se presentan en las grandes ciudades pueden llegar a ser un desafío y un problema para continuar con la dinámica regular de los flujos de personas y actividades económicas. Basta recordar lo que sucede en los pasos a desnivel, en las avenidas principales, así como en las partes bajas y marginadas de las ciudades cuando tenemos precipitaciones intensas. Desafortunadamente, el crecimiento acelerado y desorganizado de las metrópolis en México, no ha permitido cumplir las planeaciones de urbanización (Sánchez-Rodríguez et al., 2013), situación que es evidente especialmente durante condiciones de abundante lluvia.

La temporada de lluvias en Tlaquepaque, así como en el resto del Área Metropolitana de Guadalajara (AMG) inicia a mediados de junio y termina a mediados de octubre. Esto lo podemos comprobar a partir de los datos históricos de precipitación de estaciones meteorológicas, pero también, desde de la experiencia de cada uno de los habitantes del AMG. Sin embargo, pocos estudios se han realizado sobre la dinámica y variabilidad de las lluvias. García-Concepción et al., (2007) hace una estadística sobre las tormentas eléctricas locales severas en el AMG; sin embargo, no hay trabajos actualizados y/o con una metodología robusta sobre la variabilidad de la precipitación a escala estacional e interanual, así como sobre el inicio de la temporada de lluvias.

En este trabajo, se expone de manera específica y con cierto nivel de detalle, el comportamiento que han tenido las precipitaciones en el municipio de Tlaquepaque, Jalisco, desde 1951 y hasta 2022. Las preguntas que se pretenden responder, al menos en primera aproximación, están relacionadas con cantidades, periodos y fechas elementales qué deberían tenerse en cuenta en el planeamiento estratégico de crecimiento y gestión de la ciudad. Preguntas como: ¿cuánto llueve anualmente en Tlaquepaque?, ¿qué meses son los más lluviosos? ¿cuándo inicia y termina la temporada de lluvias?, ¿cuánta variabilidad tienen las precipitaciones?, ¿las lluvias qué tan extremas suelen ser?, ¿llueve más en Tlaquepaque que en el resto de los municipios del AMG? y una pregunta que puede resultar de mucho interés, ¿hay evidencia de cambio climático en cuanto a precipitaciones en Tlaquepaque? se contestarán en este artículo. Para responder a las preguntas anteriores, se realiza un procesamiento estadístico de datos horarios desde el 1 enero de 1951 y hasta el 31 diciembre de 2022 obtenidos del *reanálisis*[1] ERA-5 del *European Center for Medium-Range Weather Forecast* (ECMWF, 2023), con resolución horizontal de 0.25° × 0.25°. Por lo anterior, el objetivo principal es realizar una descripción del comportamiento, variabilidad y cambios de la temporada de lluvias, identificando aquellos aspectos de mayor interés para la planificación estratégica del presente y futuro de Tlaquepaque.

Este trabajo está organizado de la siguiente manera: en la sección 2 se describen las características del comportamiento histórico de la precipitación en Tlaquepaque, así como una breve recapitulación de algunos eventos sobresalientes; en la sección 3 se presentan evidencias y discusiones de cambio climático y en la sección 4 se concluye con algunas reflexiones.

## 2. Comportamiento histórico de la precipitación en Tlaquepaque

Conocer la climatología[2] de un lugar implica explorar lo que ha sucedió en al menos 30 años de datos. Existen diferentes fuentes de datos y técnicas estadísticas para obtener y caracterizar la climatología; sin embargo, todas ellas deben incluir primordialmente los valores promedios y medidas de dispersión de los parámetros estudiados. En esta sección, nos enfocamos en describir la climatología de 71 años de la precipitación en Tlaquepaque, desde 1951 hasta 2022, es decir, una climatología de 71 años.

### a. Descripción de la temporada de lluvias

Según la climatología calculada, la precipitación media anual en Tlaquepaque (promediando los diferentes puntos del municipio) en los últimos 71 años es de 844 ± 54 mm (Fig. 1a), muy similar al resto de los municipios que conforman el AMG, y únicamente superada por El Salto con 868 844 ± 52 mm. En la Figura 1a se observa que, los mayores valores anuales de precipitación acumulada se presentan en la mitad occidente de Jalisco, con máximos puntuales alrededor de 1400 mm en las zonas montañosas de las regiones Costa y Sur del estado. Por otro lado, hacia las regiones Norte, Altos Norte y Ciénega, los valores son menores a 800 mm con mínimos alrededor de 500 mm en el extremo nororiental de Altos Norte.

Basándose en los estudios de Stensrud et al. (1995) y Bravo-Cabrera et al. (2017), así como en la definición de inicio de temporada de lluvias del Instituto de Astronomía y Meteorología (IAM), el establecimiento de los vientos alisios (vientos del este) en combinación con el aporte de humedad de los océanos Pacífico y Atlántico, así como el paso de ondas tropicales del este, propician que el inicio de la temporada de lluvias se de a mediados del mes de junio y se extienda hasta la primera decena de octubre. No obstante, estas fechas de inicio y fin de la temporada presentan una considerable irregularidad de un año a otro y también entre las diferentes regiones de Jalisco. Por poner un ejemplo, la temporada de lluvias suele comenzar primero en la región Sur, mientras que la última región en recibir lluvias de manera sistemática es la Costa. Las lluvias de temporada en Jalisco son principalmente de tipo convectivo, es decir, se producen a partir de nubes de desarrollo vertical, que crecen durante el mediodía y la tarde precipitando en horas de la tarde-noche y algunas ocasiones extendiéndose hacia la madrugada.

---

[1] Un reanálisis es una reconstrucción meteorológica del pasado, a partir de observaciones de diferentes fuentes y datos meteorológicos modelados.

[2] La climatología de un lugar indica como se ha comportado el tiempo atmosférico, es decir, los valores medios y su variabilidad. Algunas de las variables estudiadas son precipitación, temperatura y humedad.

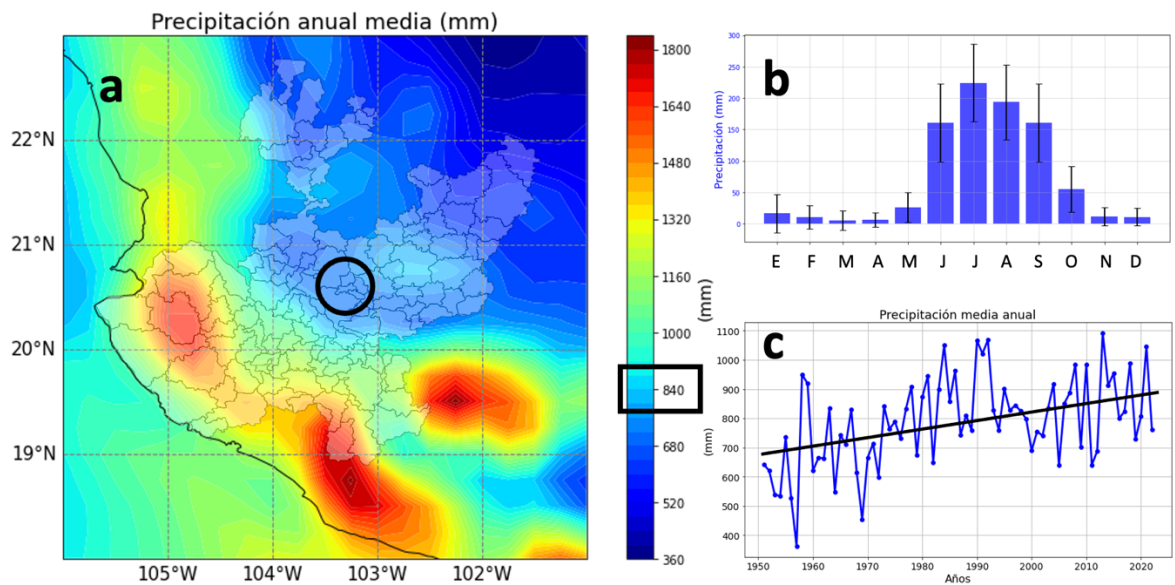

*Figura 1. (a) Precipitación media anual en Jalisco en el periodo de 1951-2022, (b) Valores acumulados medios de precipitación mensuales y su dispersión (desviación estándar) y (c) Valores medios de precipitación anuales desde en el periodo 1951-2022 y línea de tendencia.*

Como se observa en la Figura 1b, los meses con mayores registros de precipitación acumulada en Tlaquepaque son los comprendidos entre junio y septiembre. Aunque la primera quincena de octubre suele registrar prácticamente la totalidad de la precipitación del mes; es por ello por lo que el fin de la temporada se da justo en este mes. Julio es el mes más lluvioso, sin embargo, la segunda quincena de junio suele ser el periodo más lluvioso. Dicho de otro modo, aunque julio sea el mes que llueve más, la segunda quincena de junio registra la mayor precipitación acumulada. Esta es una característica muy significativa en esta región, ya que, el inicio de la temporada de lluvias es muy intenso. Teniendo en cuenta la climatología de García et al., (2007), junio es el mes con mayor número de Tormentas Eléctricas Locales Severas (TELS) y, por lo tanto, muy probablemente sea el mes con mayores daños y afectaciones a la población. Por otro lado, entre los meses de noviembre y mayo (Fig. 1b), las lluvias son muy escasas (menores al 10% del total anual); sin embargo, en ciertos años, específicamente, en años en que está presente la fase cálida del fenómeno de El Niño la Oscilación del Sur (ENSO, por sus siglas en inglés), las lluvias en diciembre y enero suelen ser superiores a la media, pudiéndose registrar en conjunto más de 50 mm.

Como se mencionó antes, eventos como el ENSO tienen un impacto directo en la variabilidad de la precipitación en el Estado de Jalisco. Se ha documentado (Bravo-Cabrera et al., 2017) que años con el evento El Niño tienen una distribución más irregular de la lluvia. Dicho de otra manera, la precipitación acumulada es muy desigual aún dentro de las mismas regiones climáticas. Por otro lado, durante el evento de La Niña (fase fría del ENSO), las lluvias presentan más uniformidad en verano, mientras que en invierno son menores de lo normal. Parte de esa variabilidad se observa en la Figura 1c; algunos años son más o menos lluviosos de lo normal, por ejemplo, entre 1980 y 1990 se observa un máximo en los valores de precipitación acumulada, mientras que entre 1995 y 2005 se presenta un mínimo. Otro aspecto en la evolución de la precipitación es su tendencia, durante el periodo de estudio, se identifica de manera muy evidente un incremento en la precipitación anual, es decir, cada vez llueve más (Fig. 1c).

### b. Sobre la variabilidad mensual en la precipitación

Dentro de los meses que conforman la temporada de lluvias en Tlaquepaque, cada uno de ellos tiene características propias, así como variabilidad interna. Considerando la climatología obtenida, se expone una descripción de los aspectos más destacados de cada mes, tales como, valores de precipitación acumulada y tendencias (Figs. 2a-e), distribuciones de precipitación acumulada (Figs. 1f-i), eventos extremos[3] y características notables.

- Junio: Es el primer mes de temporada de lluvias, con base en la definición de inicio de temporada de lluvias del Instituto de Astronomía y Meteorología de la Universidad de Guadalajara (IAM), la temporada de lluvias comienza alrededor del 11 de junio, con una desviación estándar de 3 días; es decir, la temporada de lluvias suele comenzar entre el 8 y 14 de junio. El valor acumulado mensual es de $137 \pm 43$ mm (Fig. 2a), la distribución de precipitación acumulada tiene forma de distribución Normal (o gaussiana) con eventos extremos de lluvia superiores a 250 mm e inferiores a 50 mm (Fig. 2f). Además, teniendo en cuenta que la gran mayoría de la lluvia de este mes es convectiva, es decir, se presenta en forma de tormentas eléctricas y algunas de ellas severas, los eventos de precipitación acompañados con presencia de granizo suelen estar presentes en este mes.

- Julio: Es el mes más lluvioso con $222 \pm 46$ mm (Fig. 2b), en buena medida por la influencia de ondas tropicales del este que aportan inestabilidad en combinación con las condiciones locales de calentamiento diurno y convección. La distribución de los valores de precipitación acumulada se aproxima a la normal (Fig. 2g). Eventos extremos con precipitación acumulada superior (inferior) a 320 mm (120) se han registrado, especialmente en los periodos interanuales más lluviosos (menos lluviosos) entre 1982-1995 y 2011-2018 (1950-1970 y 1996-2010). Julio es el segundo mes con mayor frecuencia de TELS. Además, durante la fase convectiva de la Oscilación Madden-Julian[4], suelen presentarse ciclos bastante lluviosos de varios días, incluso de semanas (Perdigón-Morales et al., 2021).

---

[3] Definimos eventos extremos como aquellos con valores de precipitación que se encuentran por encima o por debajo del 90% y 10% de toda la distribución.

- Agosto: Es el tercer mes de la temporada de lluvias, la precipitación acumulada es de 178 $\pm$ 55 mm (Fig. 2c). La distribución presenta una forma bimodal, es decir, con dos máximos, alrededor de 140 y 210 mm, respectivamente (Fig. 2h). Esto indica gran variabilidad interanual, años muy lluviosos (parte derecha de la distribución) y años menos lluviosos (parte izquierda de la distribución). Entre los años más lluviosos (menos lluviosos), se encuentran los que tienen valores mayores a 250 mm (menores a 100 mm). En algunos años ocurre un déficit de precipitación como consecuencia del fortalecimiento del anticiclón de las Azores, a esta situación se le conoce como sequía de medio verano[5], también llamada *canícula* (Alcalá et al., 2010).

- Septiembre: Los valores acumulados de precipitación son de 147 $\pm$ 33 mm (Fig. 2d) y la distribución sigue una forma normal (Fig. 2i) con un valor anómalo entorno a 330 mm. Se considera un mes de septiembre extremo lluvioso (poco lluvioso) como aquel que tenga valores superiores a 220 mm (inferiores a 70 mm). En la Figura 2d se identifica un mínimo de precipitación en el periodo de 1951-1976 y un periodo de máximos después de 1998, consistente con la tendencia al alza en las precipitaciones.

    Una consideración importante, es la relativa a la mayor actividad ciclónica en el océano Pacífico nororiental (Comisión Nacional del Agua [Conagua], 2023) que se da en septiembre. Debido a la circulación atmosférica, buena parte de los ciclones tropicales que se forman en este mes tienden a recurvar hacia territorio nacional (por ejemplo, los huracanes Odile 2014 y Olaf 2021). Cuando algún ciclón tropical pasa o toca tierra cerca de Jalisco, las lluvias suelen alcanzar el centro del estado, incluido el municipio de Tlaquepaque. Es usual que, bajo la influencia de un ciclón tropical, la lluvia se presente de manera intermitente y ligera durante uno o algunos días.

---

[4] Oscilación atmosférica que se asocia con la convección tropical y que tiene un periodo de 60 a 90 días (Bridgman et al., 2006 ).
[5] La sequía de medio verano consiste en una disminución en la precipitación entre dos episodios de máximos precipitación acumulada (Magaña et al., 1999).

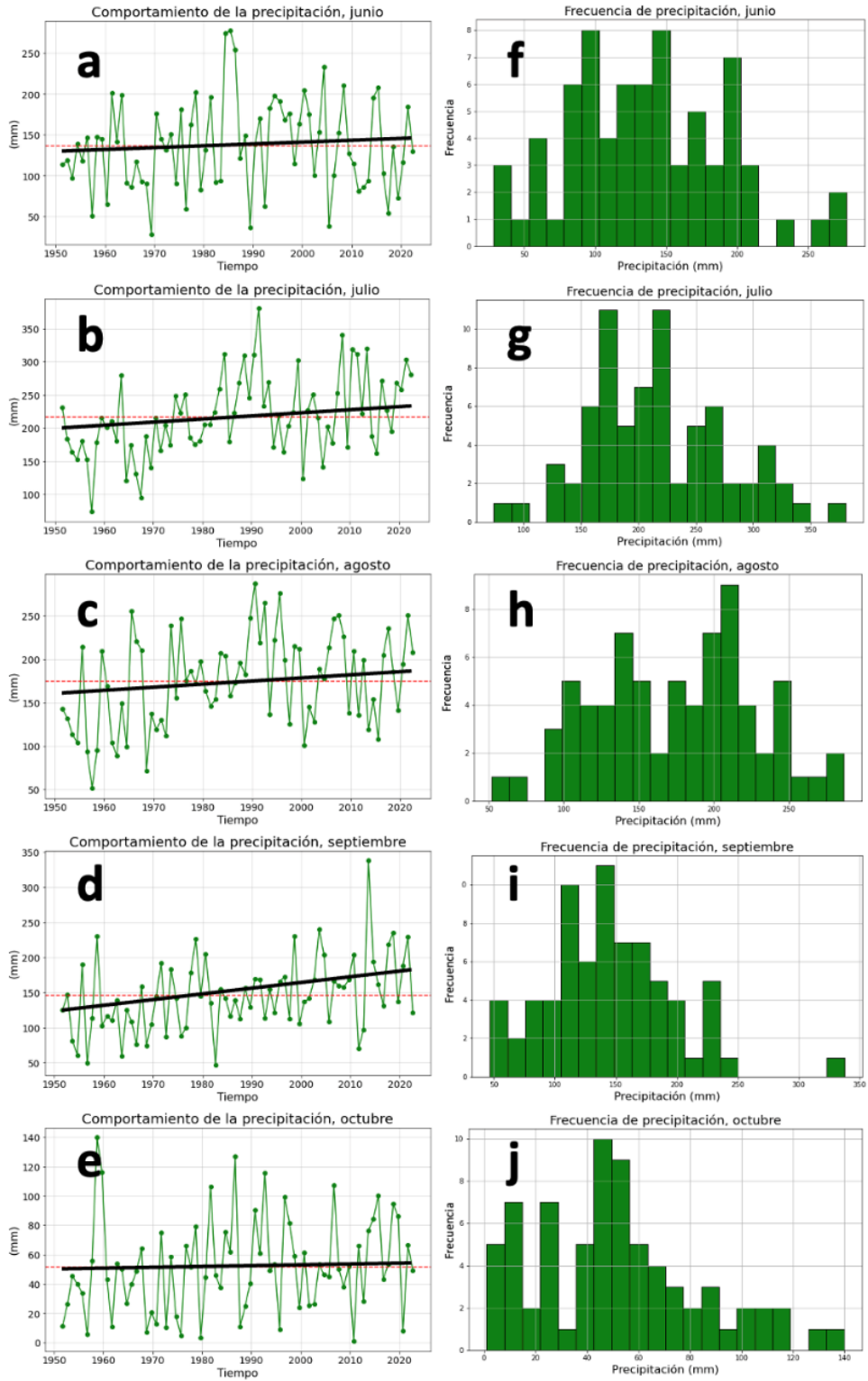

*Figura 2. (a-e) Precipitación media mensual, promedio (línea punteada roja) y tendencia, y (f-j) distribución de los valores acumulados mensuales de precipitación para los meses junio-octubre en el periodo 1951-2022.*

- Octubre: Este mes marca el fin de la temporada de lluvias. Específicamente en Tlaquepaque, las lluvias terminan alrededor del 10 de octubre con una desviación estándar de 3 días, es decir, entre el 7 y el 13 de octubre. La precipitación acumulada es de 53 $\pm$ 31 mm. La distribución de la precipitación tiene una forma normal sesgada a la izquierda con eventos extremos lluviosos (poco lluviosos) superiores a 110 mm (inferiores a 15 mm). Al comparar la precipitación media de cada mes con su desviación estándar, octubre es el mes que presenta mayor variabilidad. Esto se puede justificar por la propia variabilidad de los ciclones tropicales cerca de las costas de Jalisco, ya que octubre es el mes en que se han tenido mayor cantidad de impactos directos, incluso de huracanes de gran intensidad (López Reyes et al., 2021). Lo anterior influye en que, octubres con impactos directos o ciclones tropicales cercanos a las costas de Jalisco producen un extra de precipitación en el interior del estado.

*Tabla I. Valores mensuales de coeficientes de variación de Pearson[6].*

| Mes | Coeficiente de variación de Pearson |
|---|---|
| Junio | *0.313* |
| Julio | *0.207* |
| Agosto | *0.308* |
| Septiembre | *0.224* |
| Octubre | *0.585* |

En todos los meses se observa alta variabilidad interanual. Por ejemplo, el septiembre de 2013 se acumularon cerca de 350 mm, mientras que el septiembre del año anterior apenas se llegó a 100 mm (Fig. 2d). Esta situación se repite prácticamente en todos los meses. Sin embargo, es más evidente al inicio y final de la temporada de lluvias donde los coeficientes de variación de Pearson son mayores (Tabla I). Por lo anterior, los meses de junio y octubre son los que presentan mayor variabilidad en la precipitación acumulada.

3. **¿Hay evidencia de cambios en la precipitación en Tlaquepaque?**

Para hablar de cambio en la precipitación es necesario conocer el clima en dos periodos al menos 30 años. En el caso de este estudio, el primer periodo de estudio es entre 1951 y 1980 (llamado como clima antiguo) y el segundo entre 1993 y 2022 (llamado como clima actual). En la Figura 3 se observan las climatologías de precipitación acumulada en ambos periodos (superpuestas las distribuciones Normales o Gaussianas). Es evidente un cambio tanto en la forma como en la posición de las medias en la distribución. El clima actual se ha vuelto más lluvioso en Tlaquepaque con eventos extremos (valores acumulados superiores a 1000 mm) más frecuentes respecto al clima antiguo. De la misma manera, la variabilidad interanual ha disminuido, mostrando una distribución más leptocúrtica (Fig. 3).

---
[6] El coeficiente de variación de Pearson es una medida que indica la variabilidad de la variable en relación con el tamaño de la media, se calcula como sigue: $CV = s/\bar{x}$, donde $s$ es la desviación estandar y $\bar{x}$ la media.

Al comparar de manera mensual los cambios en la precipitación media (parte superior de la Fig. 3) se observa que todos los meses de la temporada de lluvias son más lluviosos, especialmente julio, agosto y septiembre, este último con casi 50 mm más de lluvia en la actualidad. Siendo estrictos, para la temporada de lluvias, sólo en los meses de julio, agosto y septiembre las diferencias son estadísticamente significativas al 95% (prueba de Kolmogórov-Smirnov). Por otro lado, durante los meses que no son de temporada de lluvias, se identifica una disminución cercana a 10 mm en la precipitación acumulada en el clima actual, especialmente en diciembre y enero, ambos con 95% de confianza estadística. Por lo anterior podemos decir que las temporadas de lluvias del clima actual se han vuelto más lluviosas y las temporadas no lluviosas conllevan menor precipitación.

El aumento en las lluvias se puede justificar físicamente a partir de un principio básico sobre la humedad relativa y el contenido de vapor de agua del aire (González-Alemán et al., 2023). Al aumentar la temperatura media de la Tierra, el aire tiene mayor capacidad de almacenar vapor de agua, por lo que, en términos sencillos, implica que las nubes de tormenta pueden almacenar y precipitar más agua.

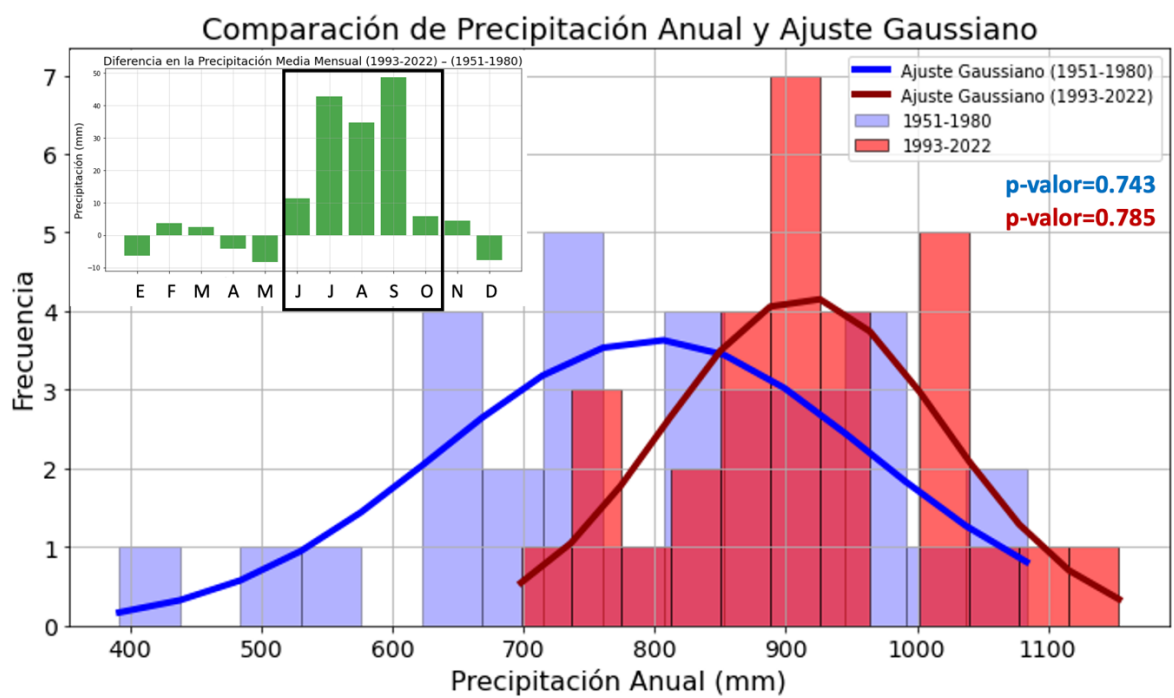

*Figura 3. Distribución de precipitación total anual para las climatologías de 1963-1992 y 1993-2022, y diferencias mensuales de la precipitación media [(1993-2022) –(1963-1992)].*

### 4. Reflexiones finales

Con el objetivo de resumir los puntos más importantes de este trabajo y con la expectativa de que esta información sirva como elemento para planificaciones futuras de gestión ambiental y territorial de Tlaquepaque en los próximos años, a continuación, se presentan algunos puntos y reflexiones finales.

A pesar de que la temporada de lluvias en Tlaquepaque comienza entre en 8 y 14 de junio, este mes presenta alta variabilidad en las precipitaciones acumuladas. Junios muy lluviosos o con un inicio adelantado de la temporada de lluvias pueden representar un reto para realización de obras públicas, actividades agrícolas y ganaderas, así como para la culminación de los trabajos de mantenimiento de sistemas de captación de aguas pluviales. Por ello, se recomiendan varias acciones:

- Anticiparse desde los primeros días de junio en las actividades previas a la temporada de lluvias.
- Vigilar el comportamiento de las condiciones y perspectivas meteorológicas desde el mes de mayo, a través de instituciones como el IAM de la Universidad de Guadalajara.

Además, junio es el mes con mayor frecuencia de TELS, por lo que, tras varios meses con escasas precipitaciones, la temporada de lluvias se inicia con cierta frecuencia de manera súbita lo que conlleva mayor impacto sobre la población y posibles complicaciones a los servicios públicos.

De la misma manera, durante el resto de los meses de temporada de lluvias, las TELS suelen estar presentes, aunque con menor frecuencia. No obstante, el análisis de cambio en la precipitación ha mostrado que el clima en Tlaquepaque se ha vuelto más lluvioso y, en combinación con el aumento demográfico, la posibilidad de impactos negativos a la población urbana es mayor. Resultaría conveniente analizar la evolución de la precipitación a lo largo del s. XXI a través de proyecciones climáticas para comprobar si la lluvia tenderá a reproducir el mismo comportamiento que presenta en la actualidad o incluso se intensificará, o por el contrario continuará una tendencia como en el periodo de clima antiguo analizado.

Finalmente, se enfatiza que, aunque la temporada de lluvias en Tlaquepaque está bien identificada y caracterizada al igual que su variabilidad interanual, el pronóstico meteorológico no puede ir a más de siete días con relativa "buena" habilidad, por lo que la información aquí presentada resulta una buena aproximación del comportamiento medio de las lluvias y puede servir para la planificación a medio y largo plazo. Sin embargo, la toma de decisiones operativas a corto plazo (unos pocos días) debe tener como referencia los pronósticos meteorológicos y el asesoramiento de recursos humanos capacitados de manera cotidiana. Todo ello con el objetivo primordial de proteger la integridad de la población y minimizar los impactos negativos en las actividades económicas.

## 5. Referencias

Alcalá, J., Meulenert A., García, O., & Ramírez Sánchez, H. U. (2013). Análisis de la canícula en el estado de Jalisco y su impacto en la agricultura. Universidad de Guadalajara, Centro Universitario de Ciencias Exactas e Ingeniería, Instituto de Astronomía y Meteorología.

Bravo-Cabrera, José Luis, Azpra-Romero, Enrique, Zarraluqui-Such, Víctor, & Gay-García, Carlos. (2017). Effects of El Niño in Mexico during rainy and dry seasons: an extended treatment. Atmósfera, 30(3), 221-232. https://doi.org/10.20937/atm.2017.30.03.03

Bridgman, Howard A., Bridgman, Howard A., Oliver, John E. (2006). The Madden-Julian oscillation (MJO). The Global Climate System: Patterns, Processes, and Teleconnections. Cambridge University Press. p. 33. ISBN 9781139455732.

Comisión Nacional del Agua (23 de octubre de 2023). Información Histórica de Ciclones Tropicales. https://smn.conagua.gob.mx/es/ciclones-tropicales/informacion-historica.

García Concepción, O., Ramírez Sánchez, H. U., Alcalá Gutiérrez, J., Meulenert Peña, Á., & García Guadalupe, M. E. (2007). Climatología de las tormentas eléctricas locales severas (TELS) en la Zona Metropolitana de Guadalajara. *Investigaciones geográficas*, (63), 7-16.

González-Alemán, J. J., Insua-Costa, D., Bazile, E., González-Herrero, S., Marcello Miglietta, M., Groenemeijer, P., & Donat, M. G. (2023). Anthropogenic Warming Had a Crucial Role in Triggering the Historic and Destructive Mediterranean Derecho in Summer 2022. Bulletin of the American Meteorological Society, 104(8), E1526-E1532. https://doi.org/10.1175/BAMS-D-23-0119.1

Lopez Reyes, M., & Meulenert Peña, Ángel R. (2021). Comparación de las variables físicas que influyen en la rápida intensificación de los ciclones tropicales del Océano Pacífico nororiental durante el periodo 1970-2018. Cuadernos Geográficos, 60(2), 105–125. https://doi.org/10.30827/cuadgeo.v60i2.15474

Magaña, V., Amador, J., & Medina, S. (1999). The Midsummer Drought over Mexico and Central America. Journal Of Climate, 12(6), 1577-1588. doi: 10.1175/1520-0442(1999)012<1577:tmdoma>2.0.co;2

Perdigón-Morales, J., Romero-Centeno, R., Ordoñez, P., Nieto, R., Gimeno, L., & Barrett, B. S. (2021). Influence of the Madden-Julian oscillation on moisture transport by the caribbean low level jet during the midsummer drought in Mexico. Atmospheric Research, 248, 105243.

Sánchez Rodríguez, R., & CEPAL, N. (2013). Respuestas urbanas al cambio climático en América Latina.

Stensrud, D. J., Gall, R. L., Mullen, S. L., & Howard, K. W. (1995). Model Climatology of the Mexican Monsoon. Journal of Climate, 8(7), 1775-1794. https://doi.org/10.1175/1520-0442(1995)008<1775:MCOTMM>2.0.CO;2.
10